# Real-time Frequency Based Reduced Order Modeling of Large Power Grid

Abilash Thakallapelli, *Student Member, IEEE*, Sudipta Ghosh, *Member, IEEE,* Sukumar Kamalasadan*, Member, IEEE*

*Abstract*—Large power systems are complex and real-time modeling of the grid for electromagnetic simulation (EMT) studies is impractical. In general, there are methods that reduce large power system into an equivalent network that requires less computational resource, while preserving electromechanical (low frequency) and high frequency behavior of the original system. This can be achieved by modeling the area not of interest (external area) as a combination of Transient Stability Analysis (TSA) type phasor model equivalent and Frequency Dependent Network Equivalent (FDNE). TSA retains electromechanical behavior, whereas FDNE retains high frequency behavior of the original power system. To this effect, this paper introduces a method of developing FDNE based on an online recursive least squares (RLS) identification algorithm in z-domain, and modeling of reduced power systems as FDNE and as a combination of TSA and FDNE using real-time digital simulators.

## I. INTRODUCTION

Using electromagnetic (EMT) simulation, power system can be modeled in detail to understand the effect of transients due to switching, resonance, etc. With large penetration of renewable energy sources like wind and photo voltaic system, the effect of power electronic components on the power grid can also be studied using EMT simulation. The integration step size of EMT simulation is in micro seconds ($\mu s$). On the other hand, in TSA type simulation only dynamic components like generators, turbines, and governors are modeled in detail with fast time dynamics such as network transients are aggregated. The integration time step of TSA models is large, which makes TSA simulation faster than EMT simulation even though not very accurate.

Detailed modeling of full scale large power system in real-time EMT simulation is impractical as this requires larger computational resource. Since modern real-time simulator's run using parallel computing processors, increase network size require more number of processors as well. One way to reduce computational time and number of processors required is to model only part of the network to be studied (study area) in which transient phenomenon or the effect of power electronic devices occurs and then represent the rest of the network (external area) as an equivalent.

The external area is generally modeled as TSA type in which the network is derived as simple inductances from short circuit impedances at the terminal buses evaluated at power frequency. This representation thus ignores high frequency oscillations. To consider high frequency transient phenomenon FDNE must be considered. If external area is modeled only as FDNE, the effect of high frequency transients on generator dynamics cannot be observed, which means the electromechanical oscillations are ignored. Thus in order to cover both electromechanical (low frequency) and high frequency behavior, external area should be modeled as TSA in parallel with FDNE.

In previous works, FDNEs are formulated as frequency-dependent black-box terminal equivalents based on rational functions. These models can easily be interfaced with EMTP type simulation platforms using a lumped circuit equivalent, or by recursive convolution. More recently FDNEs are formulated using an admittance representation. Initially short circuited admittance of a network for a wide frequency range is computed using simulated time domain responses and curve fitting techniques based on weighted polynomials. Also pole relocation (vector fitting) methods are applied for fitting the rational model to the initially computed admittance over a wide frequency range ([1]-[6]). Calculation of admittance using simulated time domain responses over a wide frequency range is tedious. For example, if frequency range of interest is from 0 to 1000 Hz and step size of frequency increment is 0.1 Hz, the total frequency samples under consideration are 10001 samples. For each frequency, sample calculation of 10001 admittances based on simulated time domain responses is extremely complex. This complexity increases with sample size of frequencies, number of ports, and size of network under consideration.

In this paper a novel method of formulating FDNEs based on online recursive least square identification is introduced. In this method external area is energized with constant voltage source after all the voltage sources are short circuited, and current sources are open circuited. Subsequently by tracking input voltage and output current, FDNE is formulated in z-domain. The proposed method is a direct way of finding FDNE which reduces complexity in finding admittance matrix over a wide frequency range and decreases computational time. To further reduce computational time number of buses (nodes) in external area for TSA type simulation are reduced using Kron's node elimination method and all generating units in external are aggregated using inertial aggregation method.

The rest of the paper is organized as follows: In section II the proposed algorithm for TSA and FDNE is discussed. For this, initially Kron's node elimination method is discussed, then generators and associated controllers aggregation are discussed, then RLS identification for FDNE formulation is discussed. In section III, implementation method of the proposed architecture in OPAL-RT (HYPERSIM)® real-time is illustrated. Section IV discusses the results on a power grid using Kundur's two area test system and Section V concludes the paper.

## II. PROPOSED ALGORITHM

The proposed algorithm has two parts a) TSA and b) FDNE. The details are discussed below.

### 1) Aggregation of External area for TSA

Initially aggregated TSA model is formulated for retaining electromechanical behavior of the external system. Aggregation is done in two steps: 1) Network aggregation, 2) Generator and associated controller aggregation.

#### a) Network Aggregation

For network aggregation first admittance matrix $'Y'$ of the external area is formulated offline using line and bus data. If there is $'n'$ no of buses in the external area and we want to retain two buses (boundary bus and generator bus) in the external area remaining $'n-2'$ buses can be eliminated using $'Y_{red}'$ [7], where

$$Y_{red} = [Y_{2x2} - Y_{2xn} Y_{nxn}^{-1} Y_{nx2}] \text{ and} \quad (1)$$

$$\begin{bmatrix} I_b \\ I_g \end{bmatrix} = Y_{red} \begin{bmatrix} V_b \\ V_g \end{bmatrix} \quad (2)$$

where, subscript $'b'$ and $'g'$ stand for the boundary bus and equivalent generator bus.

#### b) Generator and Associated Controller Aggregation

Generators and associated controllers for a large system can be aggregated before modeling TSA. Method of aggregation is discussed in [8]. Thus additional details are not explained in the paper.

### 2) Frequency Dependent Network Equivalent Formulation

In this method external area is energized with constant voltage source after short circuiting all voltage sources, and open circuiting all current sources. By tracking input voltage and output current, FDNE is then formulated in z-domain using RLS. Basic principle is as follows. If $'V_b'$ is the voltage input to the boundary bus and $'I_b'$ is the current output from the boundary bus, then frequency dependent admittance can be written as

$$Y(z) = \frac{I_b(k)}{V_b(k)} \quad (3)$$

where, $k$ is the sampling interval or sampling time.

#### a) Recursive Least Square Identification

Identification of a dynamic process is performed every sample period using the process input $'u(k)'$ and the process output $'y(k)'$ at every sample $'k'$. Considering the z-domain model of an $n^{th}$ order process, this can be represented as

$$\frac{y(k)}{u(k)} = \frac{b_1 z^{-1} + b_2 z^{-2} + \dots + b_n z^{-n}}{1 + a_1 z^{-1} + a_2 z^{-2} + \dots + a_n z^{-n}} \quad (4)$$

For $'N'$ observation window length, (4) can be rewritten as

$$\begin{bmatrix} y(k) \\ y(k-1) \\ . \\ . \\ . \\ y(k-N+1) \end{bmatrix}_{(Nx1)} = [X_{Nx2n}] \begin{bmatrix} a_1 \\ . \\ . \\ a_n \\ b_1 \\ . \\ . \\ b_n \end{bmatrix}_{(2nx1)} \quad (5)$$

Equation (5) can be written in the generic form as

$$\Phi_{model(Nx1)} = X_{(Nx2n)} \Theta_{(2nx1)} \quad (6)$$

Assume that the model identified is different from measurements, then

$$\varepsilon = \Phi_{measured} - \Phi_{model} \quad (7)$$

Where, $'\varepsilon'$ is the error between the performance of the system and model, for which criteria $'J'$ can defined as

$$J = \varepsilon^t \varepsilon \quad (8)$$

By letting $dJ/d\theta = 0$, we get

$$\Theta = [X^t X]^{-1} X^t \Phi_{measured} \quad (9)$$

From (9) it can be seen in order to get the measured variable, the inverse of the state matrix should be determined. This can drastically slow down the process and some time may even not achievable. To circumvent this issues, a recursive least squares technique is used. RLS is a computational algorithm that eliminates the matrix inversion. Let $S = X^t X$, then (9) can be written as

$$\Theta = S^{-1} X^t \Phi \quad (10)$$

where $\Phi = \Phi_{measured}$

$$\Theta(k) = S^{-1}[x(k) X^t(k-1)] \begin{bmatrix} \phi(k) \\ \Phi(k-1) \end{bmatrix} \quad (11)$$

$$\Theta(k) = S^{-1}[x(k)\phi(k) + X^t(k-1)\Phi(k-1)] \quad (12)$$

using (6)

$$\Theta(k) = S^{-1}[x(k)\phi(k) + X^t(k-1)X(k-1)\Theta(k-1)] \quad (13)$$

$$\Theta(k) = S^{-1}[x(k)\phi(k) + S(k-1)\Theta(k-1)] \quad (14)$$

since, $S(k) = S(k-1) + x(k)x'(k) \quad (15)$

Substituting (15) in (14)

$$\Theta(k) = S^{-1}[x(k)\phi(k) + \{S(k) - x(k)x'(k)\}\Theta(k-1)] \quad (16)$$

$$\Theta(k) = \Theta(k-1) + [S(k-1) + x(k)x'(k)]^{-1} x(k)[\phi(k) - x'(k)\Theta(k-1)] \quad (17)$$

Let $P(k) = S^{-1}(k)$, and by matrix inversion lemma $P(k)$ can be

represented as

$$P(k) = P(k-1)[I - \frac{x(k)x'(k)P(k-1)}{1+x'(k)P(k-1)x(k)}] \quad (18)$$

Substituting (18) in (17) and letting

$$K(k) = \frac{P(k-1)x(k)}{1+x'(k)P(k-1)x(k)} \quad (19)$$

Where P (k) can be written as

$$P(k) = [I - K(k)x'(k)]P(k-1) \quad (20)$$

Therefore, (17) can be represented as

$$\Theta(k) = \Theta(k-1) + K(k)[\phi(k) - x'(k)\Theta(k-1)] \quad (21)$$

With weighted least square, (19) and (20) can be presented as

$$K(k) = \frac{P(k-1)x(k)}{\gamma + x'(k)P(k-1)x(k)} \quad (22)$$

$$P(k) = \frac{[I - K(k)x'(k)]P(k-1)}{\gamma} \quad (23)$$

where $'\gamma'$ is the weighting factor.

This with a given process input $'V_b(k)'$ and process output $'I_b(k)'$, Y (z) can be computed using RLS identification [9].

### III. IMPLEMENTATION OF TSA AND FDNE IN HYPERSIM REAL-TIME DIGITAL SIMULATOR

Implementation method of the proposed algorithm in real-time digital simulator is a two-step process: 1) Interfacing TSA type modeling with real-time simulator, and 2) Interfacing FDNE with real-time simulator. The overall flowchart is as shown in Fig. 1.

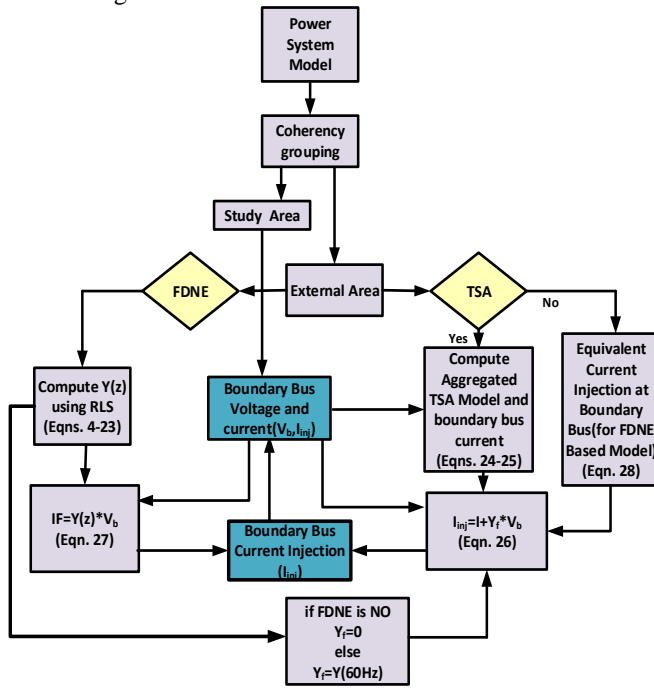

Fig 1. Implementation of FDNE and TSA.

#### 1) Interfacing TSA type modeling with real-time simulator

In this step, input to the TSA block is considered the voltage at the boundary bus, and output from the TSA block is current that is to be injected into the boundary bus. Here, first, the generator bus voltage is calculated as shown in (24). In this case, generator is modeled in detail to observe the electromechanical behavior. Conversion of boundary bus voltage from time domain to phasor domain is then performed using discrete measurement positive sequence fundamental component of the voltage. This gives magnitude $|V_b|$ and phase angle $\angle V_b$. Also phase angle of $'V_b'$ with reference to $'I_b'$ is $\angle I_b - \angle V_b$. From this,

$$V_g = (I_g - Y_{bb}V_b)Y_{gg}^{-1} \quad (24)$$

where, $I_g$ is the generator current injection and $V_b$ boundary bus voltage. Generator voltage $V_g$ is calculated recursively every time step. From calculated $V_g$ and boundary bus voltage $V_b$, boundary bus current injection $I_{binj}$ is calculated as shown in (25)-(26).

$$I_b = Y_{bb}V_b + Y_{bg}V_g \quad (25)$$

$$I_{binj} = I_b + Y(60Hz)V_b \quad (26)$$

where, $Y(60Hz)$ is the fundamental frequency component of $Y(z)$ as indicated in (3). Then $Y(60Hz)V_b$ term is added before injecting boundary bus current to remove fundamental frequency component from FDNE. After calculating boundary bus current in phasor form, the current is converted into instantaneous value and injected into the boundary bus. The overall representation is as shown in Fig. 2.

#### 2) Interfacing FDNE with real time simulator

FDNE can be directly implemented since it is computed in z-domain. With process input as $V_b$ and process output as $I_b$, and with third order identification, (4) can be written as

$$\begin{aligned}I_F(k) = &-a_1 I_F(k-1) - a_2 I_F(k-2) - a_3 I_F(k-3) \\ &+b_1 V_b(k-1) + b_2 V_b(k-2) + b_3 V_b(k-3)\end{aligned} \quad (27)$$

where $I_F$ is current output of FDNE.

If only FDNE part is used then the TSA part in Fig. 1 is ignored and boundary current is calculated from load flow results to maintain boundary bus parameters at initial steady state [14].

$$I = \left(\frac{P_b + jQ_b}{V_b \angle \theta_b}\right)^* \quad (28)$$

where, $P_b$ and $Q_b$ are the active and reactive power flow from the boundary bus, $V_b$ and $\theta_b$ are the voltage and angle of the boundary bus.

### IV. IMPLEMENTATION TEST ON INTERCONNECTED POWER GRID

For illustration purpose Kundur's two-area test system is considered (Fig. 2). The system consists of two areas containing four 900MVA synchronous generators each connected by weak tie lines. To assess the performance of our proposed FDNE-TSA based model, at first grouping of generators is done based on slow coherency method ([10]-[13]). Based on this, then the test system is divided into study and external area.

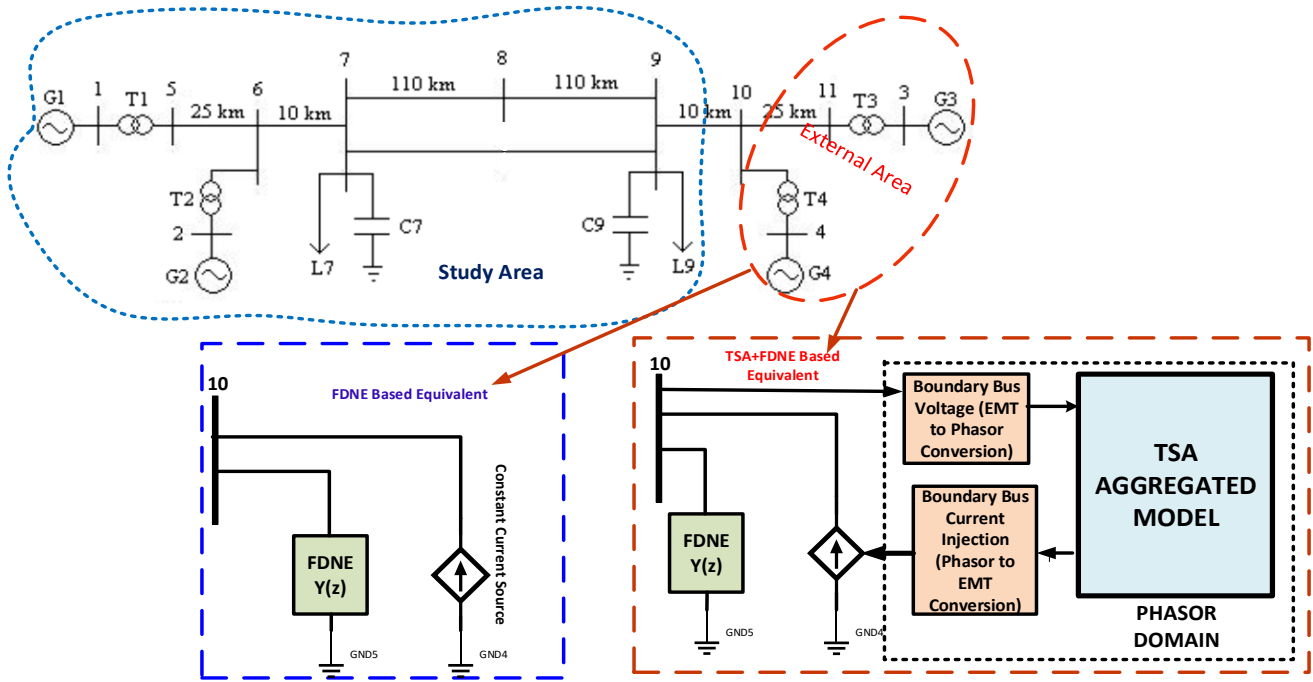

Fig. 2. Proposed dynamic equivalent of Kundur two area test system.

In this example, area-1 which includes Gen 1 & Gen 2 is considered as 'study area' and Area-2 which includes Gen 3 & Gen4 as 'external area'. Generators in external area are aggregated using inertial aggregation method and network is reduced based on Kron's reduction. Then the proposed equivalent TSA based model is developed. Finally FDNE is formulated using RLS identification procedure. Two cases are studied to evaluate the performance of the architecture. In the first case both FDNE+TSA based model is interfaced with study area at boundary bus (bus 10). In the second case TSA part is ignored and only FDNE is interfaced with the study area at boundary bus, and boundary bus is energized with constant current source calculated as shown in (28). To compare the performance of the different models a three phase short circuit fault at bus 8 was initiated at 0.1 sec and cleared at 0.2 sec (Fig. 2). Table I shows the total time required to compile and run real-time simulation for 5sec. Previous studies have indicated, for large power grid significant difference in simulation time and accuracy can be observed [15].

TABLE I
COMPARISON OF DIFFERENT MODELS

| Model | Accuracy | Speed |
|---|---|---|
| Full Model | Excellent | 8.88 sec |
| FDNE+TSA Based Model | Very Good | 7.61 sec |
| FDNE Based Model | Good | 7.5 sec |
| TSA Based Model | Poor | 7.55 sec |

Fig. 3 shows voltage at the fault bus. Fig. 4 shows the relative speed of Generator 2 with respect to Generator 1 (Swing). Fig. 5 and Fig. 6 show active power generation of Generator 1 and Generator 2 respectively. Fig. 7 shows active power flow from bus 9-7. As it can be observed, the system response is overall comparable and close to the full model representation with reduction in the computational speed.

Fig. 8 shows the power flow from the boundary bus (bus 10-9), where TSA+FDNE based and TSA based model (FDNE is ignored and only TSA part is interfaced with boundary bus) is compared with the full model. From Fig. 8 it is clear that TSA based model shows inaccuracy which can be further improved by adding FDNE. Fig. 9 shows the percentage error of active power flow from bus 10-9 with respect to the full model. Based on this test, Table I shows qualitative representation of the accuracy comparisons. From the above results it can be concluded that FDNE+TSA based equivalent model has close agreement in their behavior with the full model in all aspect.

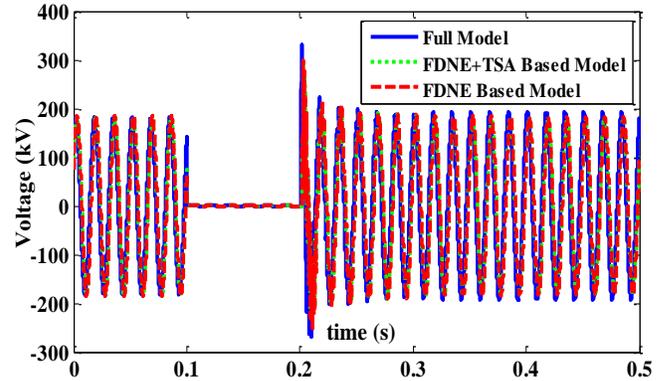

Fig. 3. The voltage at bus 8 (fault bus).

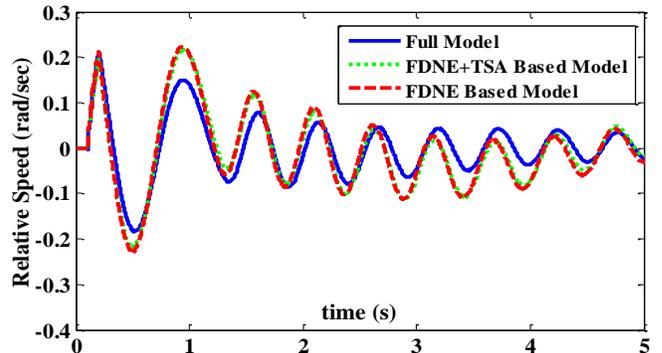

Fig. 4. Relative Speed of $G_2$ with respect to $G_1$ (swing).

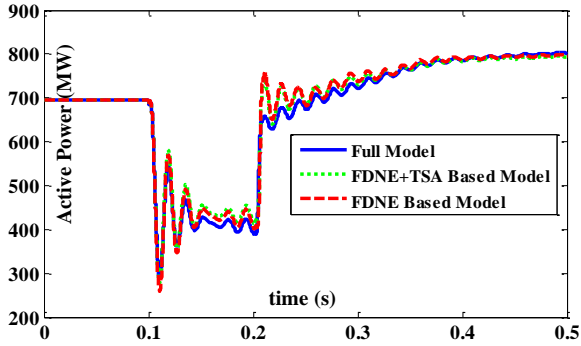
Fig. 5. The output active power of Gen 1.

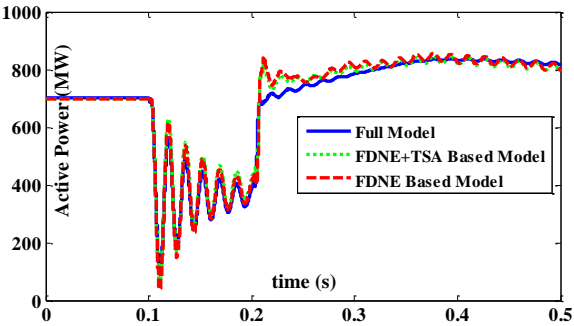
Fig. 6. The output active power of Gen 2.

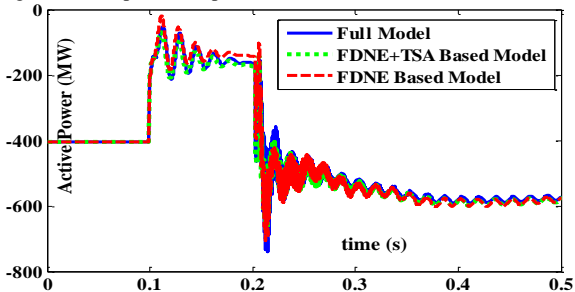
Fig. 7. The line flow from bus 9-7.

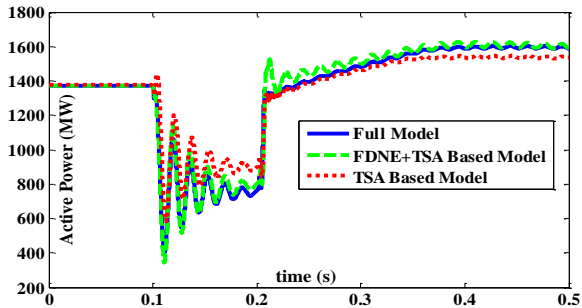
Fig. 8. Power Flow from boundary bus 10-9.

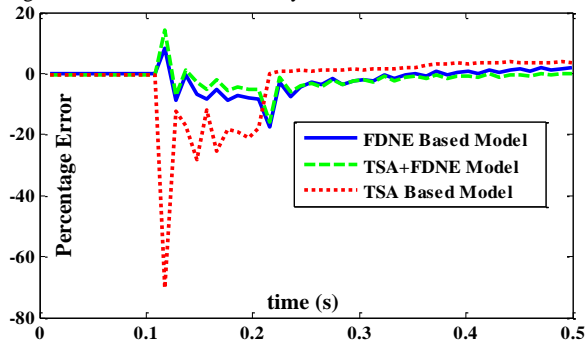
Fig. 9. Percentage error in power from bus 10-9, compared to full model.

## V. Conclusion

The proposed frequency based reduced order modeling of large power grid in the real-time simulator is simple approach which models the "study" area using a detailed EMT solution and the "external" area as equivalent. Equivalency is modeled as FDNE using RLS which represent the high-frequency behavior, and coherency based TSA to represent electromechanical behavior. The proposed method is validated on Kundur two area 4 machine systems. The reduced order model (based on FDNE+TSA) shows close agreement with full model. The advantage is that the reduced representation can replace the original model for further dynamic assessment. This proof of concept illustrates that large power systems can be modeled as equivalents in real-time simulators by reducing number of buses, and by aggregating generators and associated controllers, while retaining low frequency and high frequency behavior of the original system.